\documentclass[%
 aip,
 amsmath,amssymb,
preprint,%
]{revtex4-1}

\usepackage{graphicx}
\usepackage{dcolumn}
\usepackage{bm}
\usepackage{hyperref}

\usepackage[utf8]{inputenc}
\usepackage[T1]{fontenc}
\usepackage{mathptmx}

\begin{document}

\preprint{AIP/123-QED}

\title{Modelling the second wave of COVID-19 infections in France and Italy via a Stochastic SEIR model}

\author{Davide Faranda}
\altaffiliation{Correspondence to davide.faranda@lsce.ipsl.fr}
\affiliation{Laboratoire des Sciences du Climat et de l'Environnement,  CEA Saclay l'Orme des Merisiers, UMR 8212 CEA-CNRS-UVSQ, Universit\'e Paris-Saclay \& IPSL, 91191, Gif-sur-Yvette, France}
\affiliation{London Mathematical Laboratory, 8 Margravine Gardens, London, W6 8RH, UK}
\affiliation{LMD/IPSL, Ecole Normale Superieure, PSL research University, 75005, Paris, France}

\author{Tommaso Alberti}
\affiliation{INAF - Istituto di Astrofisica e Planetologia Spaziali, via del Fosso del Cavaliere 100, 00133 Roma, Italy}

\date{\today}

\begin{abstract}
 COVID-19 has forced quarantine measures in several countries across the world. These measures have proven to be effective in significantly reducing the prevalence of the virus. To date, no effective treatment or vaccine is available. In the effort of preserving both public health as well as the economical and social textures, France and Italy governments have partially released lockdown measures. Here we extrapolate the long-term behavior of the epidemics in both countries using a Susceptible-Exposed-Infected-Recovered (SEIR) model where parameters are stochastically perturbed to handle the uncertainty in the estimates of COVID-19 prevalence. Our results suggest that  uncertainties in both parameters and initial conditions rapidly propagate in the model and can result in different outcomes of the epidemics leading or not to a second wave of infections. Using actual knowledge, asymptotic estimates of COVID-19 prevalence can fluctuate of order of ten millions units in both countries. 
\end{abstract}

\maketitle

\section{Lead Paragraph}
\textbf{COVID-19 pandemic poses serious threats to public health as well as economic and social stability of many countries. A real time extrapolation of the evolution of COVID-19 epidemics is challenging both for the nonlinearities undermining the dynamics and the ignorance of the initial conditions, i.e., the number of actual infected individuals. 
Here we focus on France and Italy, which have partially released  initial lockdown measures. The goal is to explore sensitivity of COVID-19 epidemic evolution to the release of lockdown measures using dynamical (Susceptible-Exposed-Infected-Recovered) stochastic models. We show that the large uncertainties arising from both poor data quality and inadequate estimations of model parameters (incubation, infection and recovery rates) propagate to long term extrapolations of infections counts. Nonetheless, distinct scenarios can be clearly identified, showing either a second wave or a quasi-linear increase of total infections.}

\section{Introduction}

SARS-CoV-2 is a zoonotic virus of the coronavirus family~\cite{gaunt2010epidemiology} emerged in Wuhan (China) at the end of 2019~\cite{wu_cai_watkins_glanz} and rapidly propagated across the world until it has been declared a pandemic by the World Health Organization on March 11, 2020~\cite{world2020coronavirus}. SARS-CoV-2 virus provokes an infectious disease known as COVID-19 that has an incredibly large spectrum of symptoms or none depending on the age, health status and the immune defenses of each individuals~\cite{covid2020severe}. SARS-CoV-2 causes potentially life-threatening form of pneumonia and/or cardiac injuries in a non-negligible patients fraction~\cite{zheng2020covid,huang2020clinical}. 
 
To date, no treatment of vaccine is available for COVID-19~\cite{cascella2020features}. Efforts to contain the virus and to not overwhelm intensive care facilities are based on quarantine measures which have proven very effective in several countries~\cite{anderson2020will, Chinazzi20, Yuan20}. Despite this, lockdown measures entail enormous economical, social and psychological costs. Recent estimates of the International Monetary Fund recently announced a global recession that will drag global GDP lower by 3\% in 2020, although continuously developing and changing as well as significantly depending country-by-country~\cite{fernandes2020economic}. More than 20 million people have lost their job in United States~\cite{coibion2020labor} and a large percentage of Italians have developed psychological disturbances such as insomnia or anxiety due to the strict lockdown measures~\cite{cellini2020changes}. Those measures have been taken on the basis of epidemics models, which are fitted on the available data~\cite{rothan2020epidemiology}. In Italy, initial lockdown measures started on February 23rd for 11 municipalities in both Lombardia and Veneto which were identified as the two main Italian clusters. After the initial spread of the epidemics into different regions all Italian territory was placed into a quarantine on March 9th, with total lockdown measures including all commercial activities (apart supermarkets and pharmacies), non-essential businesses and industries, and severe restrictions to transports and movements of people at regional, national, and extra-national levels\cite{chintalapudi2020covid}. People were asked to stay at home or near for sporting activities and dog hygiene (within 200 m from home), to reduce as much as possible their movements (only for food shopping and care reasons), and smart-working was especially encouraged in both public and private administrations and companies. At the early stages of epidemics intensive cares were almost saturated with a peak of $~$4000 people on April 3rd and a peak of hospitalisations of $~$30000 on April 4th, significantly reducing after these dates, reaching $~$1500 and $~$17000, respectively, at the beginning of phase 2 on May 4th, and $~$750 and $~$1000 on May 18th when lockdown measures on commercial activities were relaxed. These numbers, continuously declining during the next days and weeks, confirmed the benefit of lockdown measures \cite{Gatto20}.\\
Alarmed by the exponential growth of new infections and the saturation of the intensive care beds, also France introduced strict lockdown measures on March 17th~\cite{roux2020covid}. The French government restricted travels to food shopping, care and work when teleworking was not possible, outings near home for individual sporting activity and/or dog hygiene, and it imposed the closure of the Schengen area borders as well as the postponement of the second round of municipal elections. The number of patients in intensive care, like the number of hospitalisations overall peaked in early April and then started to decline, showing the benefits of lockdown measures. On Monday, May 11th, France began a gradual easing of COVID-19 lockdown measures~\cite{di2020expected}. Trips of up to 100 kilometres from home are allowed without justification, as will gatherings of up to 10 people. Longer trips will still be allowed only for work or for compelling family reasons, as justified by a signed form. Guiding the government’s plans for easing the lockdown is the division of the country into two zones, green and red, based on health indicators. Paris region (Ile de France), with about 12 milions inhabitants is flagged, to date, as an orange zone.
 
In both countries, the release of lockdown measures has been authorised by authorities after consulting scientific committees which were monitoring the behavior of the curve of infections using COVID-19 data. Those data are provided daily, following a request of the WHO. To date, the WHO guidelines require countries to report, at each day $t$, the total number of infected patients $I(t)$ as well as the number of deaths $D(t)$. Large  uncertainties have been documented in the count of $I(t)$~\cite{ghoshal2020estimating}. Whereas in the early stage of the epidemic several countries tested asymptomatic individuals to track back the infection chain, recent policies to estimate $I(t)$ have changed. Most of the western countries have previously  tested only patients displaying severe SARS-CoV-2 symptoms~\cite{hale2020variation}. In an effort of tracking all the chain of infections, Italy and France are now testing all individuals displaying COVID-19 symptoms and those who had strict contacts with infected individuals.  The importance of tracking asymptomatic patients has been proven in a recent study~\cite{li2020substantial}. The authors have estimated that an enormous part of total infections were undocumented  (80\% to 90\%)  and that those undetected infections were the source for 79\% of documented cases in China. Tracking strategies have proven effective in supporting actions to reduce the rate of new infections, without the need of lockdown measures, as in South Korea~\cite{nunes2020visualising}. 

The goal of this paper is to explore possible future epidemics scenarios of the long term behavior of the COVID-19 epidemic~\cite{desai2019real} but taking into account the role of uncertainties in both the parameters value and the infection counts to investigate different outcomes of the epidemics leading or not to a second wave of infections. To this purpose we use a stochastic Susceptible-Exposed-Infected-Recovered (SEIR) model~\cite{Faranda20} which consist in a set of ordinary differential equations where control parameters are time-dependent modelled via a stochastic process. This allows to mimic the dependence on control parameters on some additional/external factors as super-spreaders \cite{lloyd2005superspreading} and the enforcing/relaxing of confinement measures \cite{Faranda20}. As for the classical SEIR models \cite{brauer2008compartmental} the population is divided into four compartmental groups, i.e., Susceptible, Exposed, Infected, and Recovered individuals. The stochastic SEIR model shows that long-term extrapolation is sensitive to both the initial conditions and the value of control parameters \cite{Faranda20}, with asymptotic estimates fluctuating on the order of ten millions units in both countries, leading or not a second wave of infections. This sensitivity arising from both poor data quality and inadequate estimations of model parameters has been also recently investigated by means of a statistical model based on a generalized logistic distribution \cite{Alberti20}. 
The paper is organised as follows: in Section \ref{sect:datamodelling} we discuss the various sources of data for COVID-19 and their shortcomings, and then we discuss in detail the SEIR model and its statistical modelling. In Section \ref{sect:results} we discuss the results focusing on the statistical sensitivity of the modelling, and apply it to data from France and Italy. We finish, in Section \ref{sec:discussion}, with some remarks and point out some limitiations of our study.

\section{Data and modelling}
\label{sect:datamodelling}
\subsection{Data}
This paper relies on data stored into the Visual Dashboard repository of the Johns Hopkins University Center for Systems Science and Engineering (JHU CSSE) supported by ESRI Living Atlas Team and the Johns Hopkins University Applied Physics Lab (JHU APL). Data can be freely accessed and downloaded at \url{https://systems.jhu.edu/research/public-health/ncov/}, and refers to the confirmed cases by means of a laboratory test~\cite{world2020coronavirus}. Nevertheless there are some inconsistencies between countries due to different protocols in testing patients (suspected symptoms, tracing-back procedures, wide range tests) \cite{d'emilio_winfield,arin}, as well as, to local management of health infrastructures and institutions. As an example due to the regional-level system of Italian healthcare data are collected at a regional level and then reported to the National level via the Protezione Civile transferring them to WHO. These processes could be affected by some inconsistencies and delays~\cite{agi}, especially during the most critical phase of the epidemic diffusion that could introduce errors and biases into the daily data. These incongruities mostly affected the period between February 23rd and March 10th, particularly regarding the counts of deaths due to a protocol change from the Italian Ministry of Health \cite{Ferrari20}. A similar situation occurs in France where the initial testing strategy was based only on detecting those individuals experiencing severe COVID19 symptoms~\cite{cohen2020countries}. In the post lockdown phase, France has extended its testing capacity to asymptomatic individuals who have been in contact with infected patients~\cite{tanne2020covid}.

\subsection{A Stochastic epidemiological Susceptible-Exposed-Infected-Recovered model}
One of the most used epidemiological models is the so-called Susceptible-Exposed-Infected-Recovered (SEIR) model belonging to the class of compartmental models~\cite{brauer2008compartmental}. It assumes that the total population $N$ can be divided into four classes of individuals that are susceptible $S$, exposed $E$, infected $I$, and recovered or dead $R$ (assumed to be not susceptible to reinfection). The model is based on the following assumptions: 
\begin{enumerate}
    \item the total population does not vary in time, e.g., $dN/dt = dS/dt + dE/dt  + dI/dt + dR/dt = 0, ~~\forall t \geq 0$;
    \item susceptible individuals become infected that then can only recover or die, e.g., $S \to I \to R$;
    \item exposed individuals $E$ encountered an infected person but are not themselves infectious;
    \item recovered or died individuals $R$ are forever immune. 
\end{enumerate}
Thus, the model reads as
\begin{eqnarray}
\frac{dS}{dt} & = & - \lambda S(t) I(t), \label{eq2aSDE} \\
\frac{dE}{dt} & = &   \lambda S(t) I(t) -\alpha E(t), \label{eq2bSDE}\\
\frac{dI}{dt} & = & \alpha E(t)- \gamma I(t), \label{eq2cSDE} \\
\frac{dR}{dt} & = & \gamma I(t), \label{eq2dSDE}
\end{eqnarray}
where $\gamma > 0$ is the recovery/death rate, $\lambda = \lambda_0/S(0) > 0$ is the infection rate rescaled by the initial number of susceptible individuals $S(0)$, and $\alpha$ is the inverse of the incubation period. Its discrete version can be simply obtained via an Euler Scheme as

\begin{eqnarray}
S(t+1) & = & S(t) - \lambda S(t) I(t), \label{eq2a} \\
E(t+1) & = & (1-\alpha)E(t) +  \lambda S(t) I(t),  \label{eq2b}\\
I(t+1) & = & (1-\gamma)I(t) + \alpha E(t), \label{eq2c} \\
R(t+1) & = & R(t) + \gamma I(t) . \label{eq2d}
\end{eqnarray}

in which we fixed $dt=1$ day that is the time resolution of COVID-19 counts. By means of $\gamma$ and $\lambda_0$ the model also allows to derived the so-called $R_0$ parameter, e.g., $R_0=\lambda_0/\gamma$, representing the average reproduction number of the virus. It is related to the number of cases that can potentially (on average) caused from an infected individual during its infectious period ($\tau_{inf} = \gamma^{-1}$). Early estimates in Wuhan~\cite{wu2020nowcasting} on January 2020 reported $R_0 = 2.68_{2.47}^{2.86}$ which lead to $\gamma = 0.37$  fixing $\lambda\simeq 1$ as in \cite{peng2020epidemic} and a 95\% confidence level range for the incubation period between 2 and 11 days\cite{lauer2020incubation}. However, the $R_0$ parameter as well as models parameters $\lambda$, $\gamma$, and $\alpha$ can vary in time during the epidemics due to different factors as the possible presence of the so-called super-spreaders~\cite{lloyd2005superspreading}, intrinsic changes of the SARS-CoV-2 features, lockdown measures, asymptomatic individuals who are not tracked out, counting procedures and protocols, and so on \cite{Lavezzo20}. 

To deal with uncertainties in long-term extrapolations and with the time-dependency of control parameters a stochastic approach could provide new insights in modeling epidemics~\cite{olsen1990chaos,andersson2012stochastic,dureau2013capturing}, especially when epidemics show a wide range of spatial and temporal variability ~\cite{polonsky2019outbreak,viceconte2020COVID,kashnitsky2020COVID}. However, instead of investigating how to get a realistic behavior by stochastically perturbing control parameters, here we investigate how uncertainties into the final counts $C(t)$ are controlled by model parameters \cite{Faranda20}. Thus, we use a stochastic version of the SEIR model in which the set of control parameters $\kappa \in \{\alpha,\lambda, \gamma\}$ is modelled via a stochastic process

\begin{equation}
\kappa(t)=|\kappa_0 +\sigma\cdot\xi(t)|,
\label{eqstochpert}
\end{equation}

with being $\sigma$ the intensity of the perturbation, $\xi(t)$ a random variable from a collection of normally distributed $N(0,1)$ values at each time $t$, and since $\kappa(t) \ge 0$ we introduced the absolute value in Eq.~(\ref{eqstochpert}). In this way we can introduce instantaneous daily discrete jumps (e.g., take into account daily uncertainties) in the control parameters to properly model detection errors on infection counts, appropriately described through a discrete process~\cite{faranda2014extreme} than a continuous one~\cite{faranda2017stochastic}.  

\section{Results}
\label{sect:results}

\subsection{Model validation: first wave}

We begin this section by validating the SEIR stochastic model on the first wave of infections. We have therefore to chose the initial conditions, and then introduce the lockdown measures in the parameters.

\begin{figure}[ht]
\begin{center}
\includegraphics[width=1.08\textwidth]{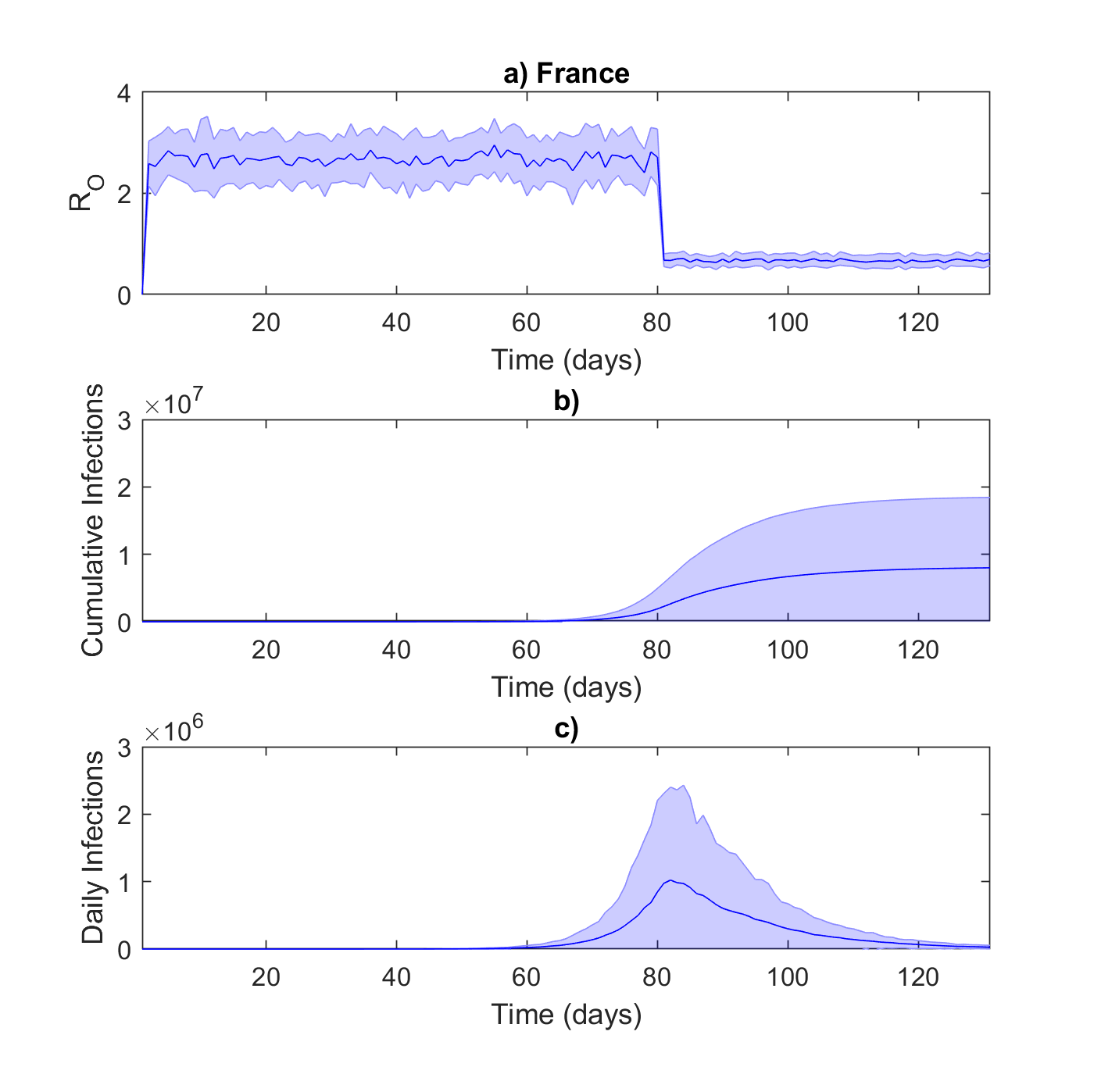}
\caption{ Susceptible-Exposed-Infected-Recovered (SEIR) model of COVID-19 for France (Eqs~\ref{eq2a}-\ref{eq2d}) with $\lambda=1./S(0)$, $\alpha=0.27$, $\gamma=0.37$. Initial conditions are set to $I(1)=1$, $S(1)=6.7\cdot  10^7$, $E(1)=R(1)=0$. $t=1$ corresponds to Dec 27, 2019. Confinement is introduced at $t=78$ (Mar 17, 2020).  a) Time evolution for the basic reproduction number $R_0$, b) Time evolution for the cumulative number of infections $C(t)$, c) Time evolution for the daily infected individuals $I(t)$. Solid line shows the average for 30 realisation of the SEIR stochatic models, shading extends to 3 standard deviations of the mean. }
\label{Fig1}
\end{center}
\end{figure}

\paragraph{France\\}
In France, the first documented case of COVID-19 infections goes back to December 27th, 2019. Doctors at a hospital in the northern suburbs of Paris retested samples from patients between December 2nd, 2019, and January 16th, 2020. Of the 14 patient samples retested, one sample, from a 42-year-old man came back positive~\cite{deslandes2020sars}. As initial condition for the SEIR model, we therefore set $I(t=1)=1$ and $t=1$ corresponds to December 27th, 2019. We then use $R_0 = 2.68_{2.47}^{2.86}$ which lead to $\gamma = 0.37$  fixing $\lambda_0\simeq 1$. Strict lockdown measures are introduced at $t=80$ (i.e., March 17th, 2020). First wave modelling results are shown in Figure~\ref{Fig1}. Figure~\ref{Fig1}a) shows the modelled value of $R_0$. During confinement, we let $\lambda$ fluctuates by 20\% of its value around 1/4. We base this new infection rate on the mobility data for France during confinement, which have shown a drop by $\sim 75\%$ according to the INSERM report \#11~\cite{pullanopopulation}. The resulting confinement $R_0=0.75_{0.5}^{1.0}$, a range of values compatible with that published by the Pasteur Institute~\cite{salje2020estimating}.  The cumulative number of infections is shown in Figure~\ref{Fig1}b) and shows, on average, that 8 millions people have been infected by SARS-CoV-2 in France. The uncertainty range is extremely large, according to the error propagation given by the stochastic fluctuations of the parameters (see \cite{Faranda20} for explanations). It extends from few hundred thousands individuals up to 18 millions. The average is however close to the value proposed by the authors in ~\cite{Salje20}, who estimate a prevalence of $\sim 6\%$ of COVID-19 in the French population. Another realistic feature of the model is the presence of an asymmetric behavior of the right tail of daily infections distributions (Figure~\ref{Fig1}c) that has also been observed in real COVID-19 published data~\cite{maleki2020time}.

\paragraph{Italy\\}

\begin{figure}[ht]
\begin{center}
\includegraphics[width=1.08\textwidth]{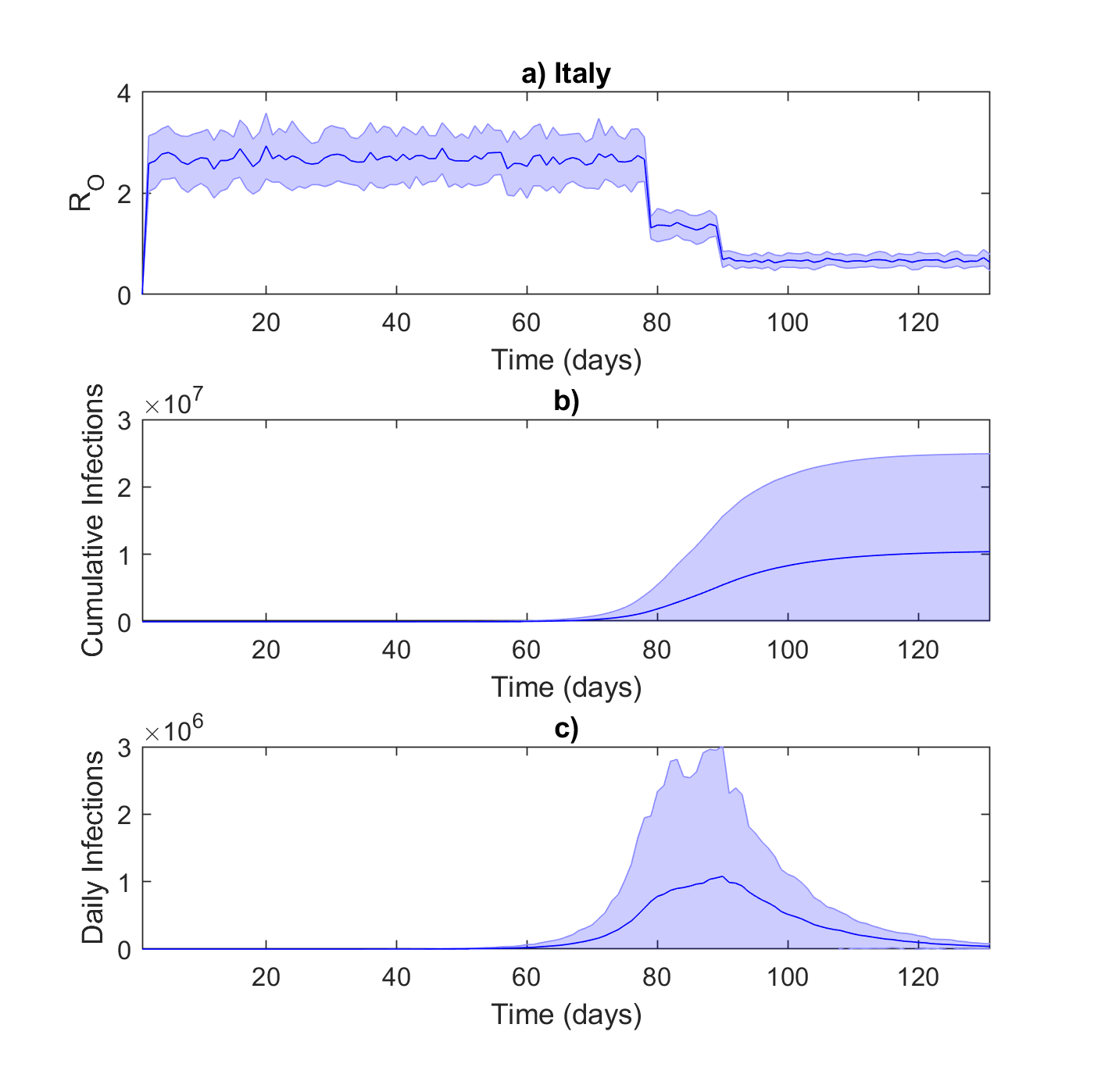}
\caption{ Susceptible-Exposed-Infected-Recovered (SEIR) model of COVID-19 for Italy (Eqs~\ref{eq2a}-\ref{eq2d}) with $\lambda=1./S(0)$, $\alpha=0.27$, $\gamma=0.37$. Initial conditions are set to $I(1)=1$, $S(1)=6.0\cdot  10^7$, $E(1)=R(1)=0$. $t=1$ corresponds to Dec 22, 2019. First confinement measures are  introduced at $t=78$ (Mar 9, 2020) and enforced at $t=89$ (Mar 22, 2020). a) Time evolution for the basic reproduction number $R_0$, b) Time evolution for the cumulative number of infections $C(t)$, c) Time evolution for the daily infected individuals $I(t)$. Solid line shows the average for 30 realisation of the SEIR stochatic models, shading extends to 3 standard deviations of the mean. }
\label{Fig2}
\end{center}
\end{figure}

For Italy, the first suspect COVID-19 case goes back to December 22nd, 2019, a 41-year-old woman who could only be tested positive for SARS-CoV-2 antibodies in April 2020~\cite{corriere_della_sera_2020primo}. As initial condition we therefore set $I(t=1)=1$ and $t=1$ corresponds to December 22nd, 2019. As for France we use $R_0 = 2.68_{2.47}^{2.86}$ leading to $\gamma = 0.37$ if fixing $\lambda_0\simeq 1$. A first semi-lockdown was set in Italy on March 9th, 2020 ($t=78$) and enforced on March 22nd, 2020 ($t=89$). To simulate these two-steps lockdown we again base our reduction in $R_0$ on the mobility data for Italy which show for the first part of the confinement a reduction of about 50~\% and a similar reduction to France (75\%) for the strict lockdown phase. Figure~\ref{Fig2} shows the results for the first wave by letting $\lambda = 0.25 \pm \Delta \lambda$, where $\Delta \lambda$ represents a 20\% fluctuations around the mean value, and by fixing an initial condition on susceptible individuals $S(1) = 6.0 \cdot 10^7$ corresponding to the estimate of the Italian population. A clear difference emerges with respect to the case of France in the behavior of $R_0$ which shows an intermediate reduction near $t=80$, corresponding to March 11th, 2020, to $R_0 = 1.4_{1.1}^{1.7}$ before reaching the final value of $R_0=0.7_{0.5}^{0.9}$. This sort of "step" into the $R_0$ time behavior corresponds to the time interval between semi- and full-lockdown measures, whose efficiency significantly increases after March 24th, 2020, also corresponding to the peak value of infections. This is confirmed by looking at daily infections distributions (Figure~\ref{Fig2}c) that shows a peak value near March 24th, 2020, also observed in real COVID-19 data \cite{Alberti20}. Finally, the cumulative number of infections (Figure~\ref{Fig2}b) shows that, on average, almost 10 millions people have been infected by SARS-CoV-2 in Italy, ranging between few hundred thousands up to 25 millions due to the the error propagation by the stochastic fluctuations of model parameters (see \cite{Faranda20} for explanations). Nevertheless the wide range of uncertainty the average value is close to the value estimated from a team of experts of the Imperial College London according to which the 9.6\% of Italian population has been infected, with a 95\% confidence level ranging between 3.2\% and 26\%~\cite{flaxman2020report}. These estimates correspond to cumulative infections of $\sim$6 millions, ranging from $\sim$2 and $\sim$16 millions, well in agreement with our model and other statistical estimates \cite{DeNatale20}. 

\subsection{Future epidemics scenarios}

After lockdown measures are released, for both countries, we model three different scenarios: a first one where all restrictions are lifted (back to normality), a second one where strict measures are taken (semi-lockdown) and a third one where the population remains mostly confined (lockdown).   
\paragraph{France\\}

\begin{figure}[ht]
\begin{center}
\includegraphics[width=0.9\textwidth]{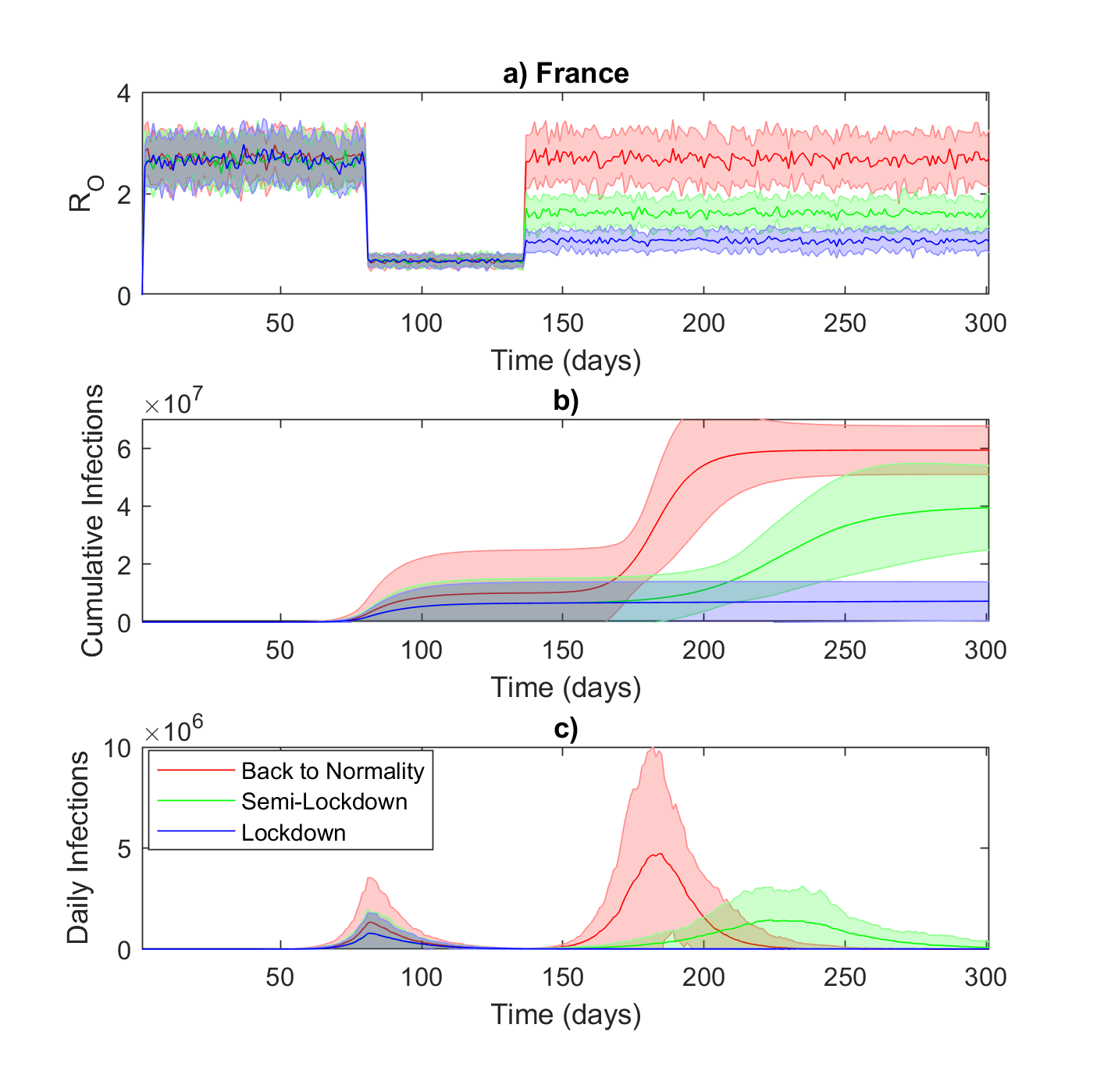}
\caption{ Susceptible-Exposed-Infected-Recovered (SEIR) model of COVID-19 for the second wave in France. Initial conditions are set as in Figure~\ref{Fig1}. After the confinement is released ($t=136$, May 11, 2020) three scenarios are modelled: back to normality (red), semi-lockdown(green), lockdown (blue).   a) Time evolution for the basic reproduction number $R_0$, b) Time evolution for the cumulative number of infections $C(t)$, c) Time evolution for the daily infected individuals $I(t)$. Solid line shows the average for 30 realisation of the SEIR stochatic models, shading extends to 3 standard deviations of the mean. }
\label{Fig3}
\end{center}
\end{figure}

Results for France are shown in Figure~\ref{Fig3}. Lockdown is released  at $t=136$, corresponding to May 11th, 2020. The  back to normality (red) scenario clearly shows a second wave of infections peaking in summer (early July) and forcing group immunity in the French population. The semi-lockdown (green) scenario, corresponding to a reduction of the mobility of about 50\%, leads to a second wave as intense as the first wave, but longer, at the end of August. As in the previous scenario, the semi-lockdown scenario allows to reach a group immunity in France. A third lockdown scenario is modelled (blue). This latter scenario simulates an $R_0\simeq 1$, that can be achieved by imposing strict distancing measures, contact tracking as well as reduction in mobility. It results in a linear modest increase of the total number of infections that does not produce a proper wave of infections. As in the first wave modelling, large uncertainties are also present in future scenarios although the three distinct behaviors clearly appear. 

\paragraph{Italy\\}

\begin{figure}[ht]
\begin{center}
\includegraphics[width=0.98\textwidth]{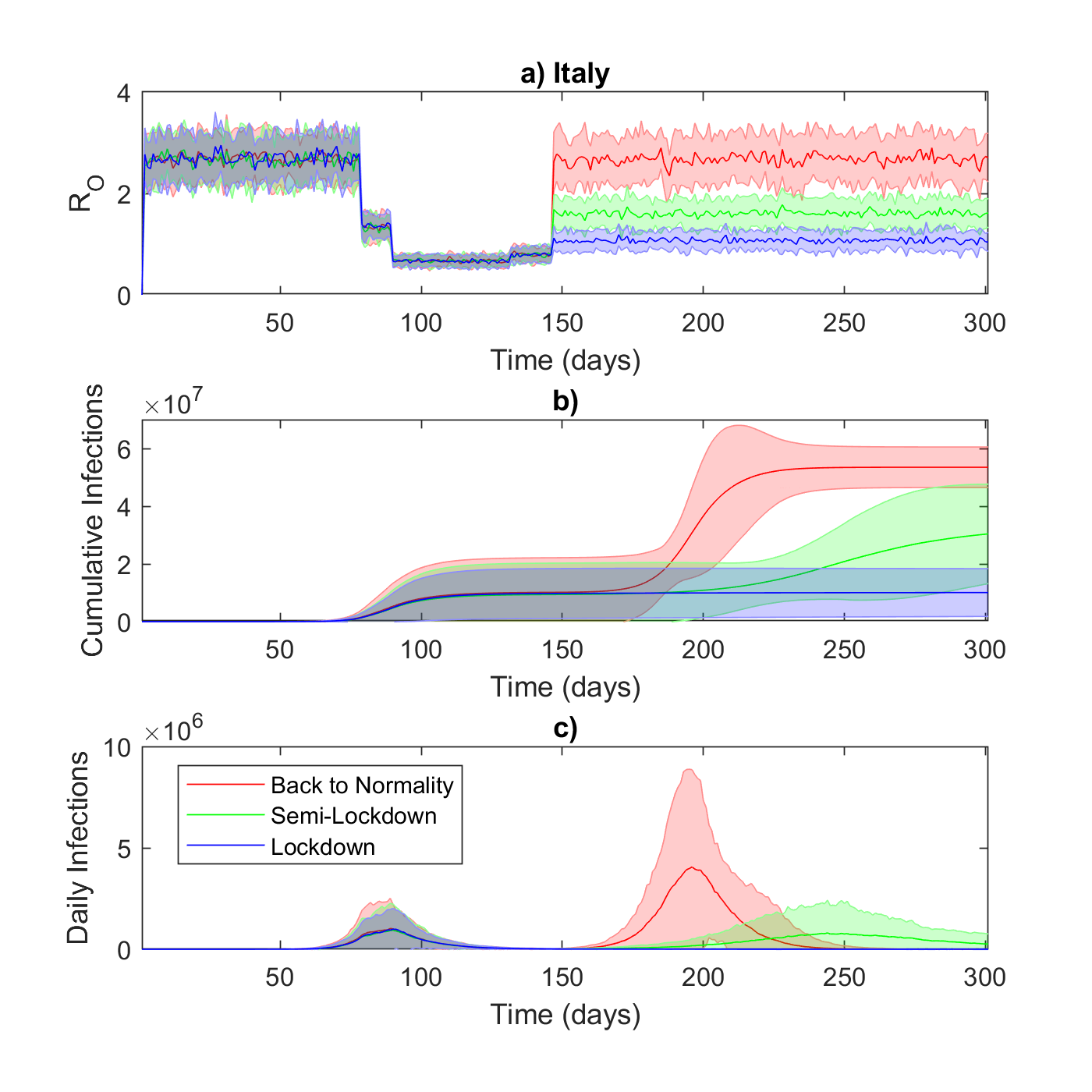}
\caption{ Susceptible-Exposed-Infected-Recovered (SEIR) model of COVID-19 for the second wave in Italy. Initial conditions are set as in Figure~\ref{Fig2}. After the confinement is released ($t=131$, May 4, 2020 and $t=146$, May 18, 2020) three scenarios are modelled: back to normality (red), semi-lockdown (green), lockdown (blue).   a) Time evolution for the basic reproduction number $R_0$, b) Time evolution for the cumulative number of infections $C(t)$, c) Time evolution for the daily infected individuals $I(t)$. Solid line shows the average for 30 realisations of the SEIR stochatic models, shading extends to 3 standard deviations of the mean. }
\label{Fig4}
\end{center}
\end{figure}

Figure \ref{Fig4} shows the results for modeling future epidemic scenarios for Italy. The first relaxation of lockdown measures started at $t=131$, corresponding to May 4th, 2020, while strict measures were finally released at $t=146$, corresponding to May 18th, 2020. The back to normality (red) scenario moves towards a second wave of infections whose peak occurs at $t=193$, corresponding to July 4th, 2020, exactly three months after initial lockdown measures were released (May 4th, 2020). This would lead the so-called herd immunity for the whole Italian population (see Figure~\ref{Fig4}b), with a peak of daily infections near 5 millions of people (Figure~\ref{Fig4}c), and $R_0$  re-approaching the initial value ($R_0 = 2.68)$. The semi-lockdown (green) scenario produces a second wave mostly similar, in terms of intensity, as the first wave, but occurring at $t=246$, e.g., August 26th, 2020. This scenario will lead to 40 millions infected people, spanning between 25 and 55 millions, thus producing a group immunity in Italy. A third scenario is modelled in which complete lockdown measures are still considered (blue). This latter scenario leads to a more controlled evolution of cumulative infections which still remain practically unchanged with respect to the first wave cumulative number. It has been obtained by simulating an $R_0\simeq 1$, resulting from strict distancing measures and reduced mobility, and does not produce a proper wave of infections. However, all scenarios are clearly characterized by a wide range of uncertainties, although producing three well distinct behaviors in both cumulative and daily infections.

\subsection{Phase Diagrams}
\begin{figure}[ht]
\begin{center}
\includegraphics[width=0.98\textwidth]{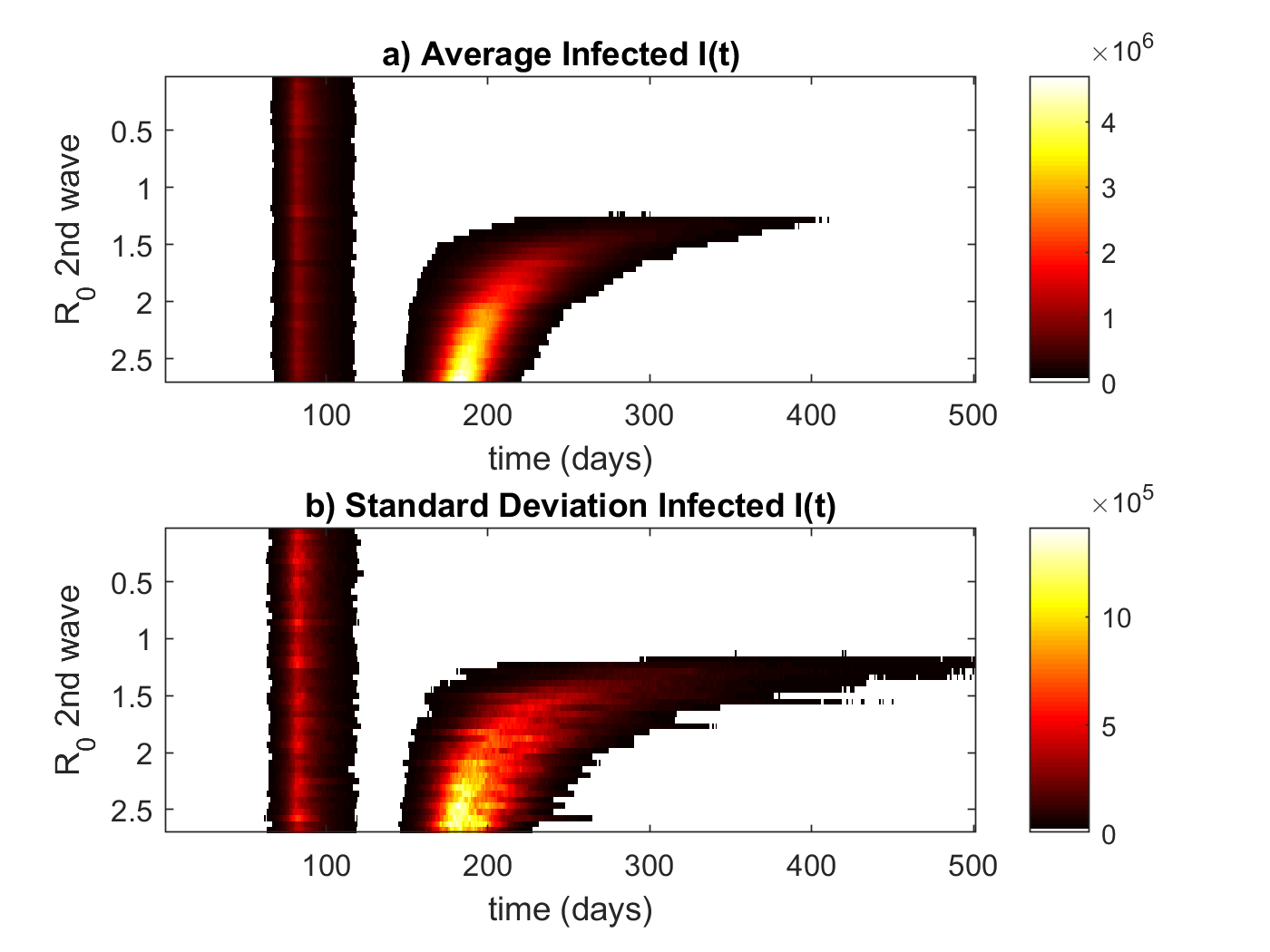}
\caption{Phase diagram for the Susceptible-Exposed-Infected-Recovered (SEIR) model of COVID-19 for the second wave in France. Initial conditions are set as in Figure~\ref{Fig1}. After the confinement is released ($t=136$, May 11, 2020) all possible $R_0$ modelled.  a) Average of daily infected individuals $I(t)$. b) Standard deviation of daily infected individuals. Diagrams are obtained using 30 realisations of the SEIR models.}
\label{Fig5}
\end{center}
\end{figure}

\begin{figure}[ht]
\begin{center}
\includegraphics[width=0.98\textwidth]{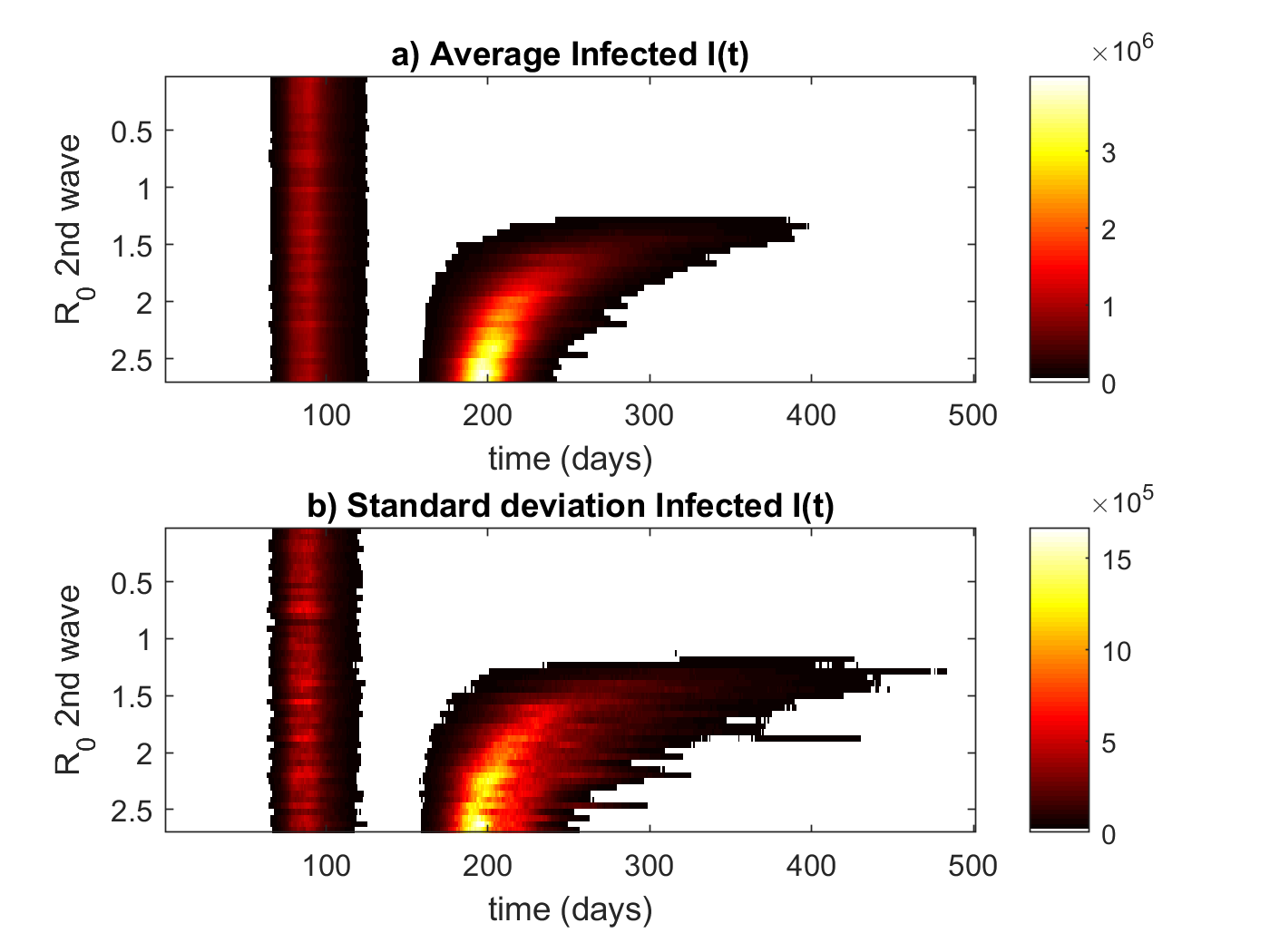}
\caption{Phase diagram for the Susceptible-Exposed-Infected-Recovered (SEIR) model of COVID-19 for the second wave in Italy. Initial conditions are set as in Figure~\ref{Fig2}. After the confinement is released ($t=131$, May 4, 2020 and then $t=146$ May 18, 2020)   all possible $R_0$ modelled.  a) Average of daily infected individuals $I(t)$. b) Standard deviation of daily infected individuals. Diagrams are obtained using 30 realisations of the SEIR models.}
\label{Fig6}
\end{center}
\end{figure}

In the previous section we have seen that increasing $R_0$ above 1 can or not produce a second wave of infections and introduce also a time delay in the appearance of a second wave of infections. We now analyse this effect in a complete phase diagram fashion. Figures~\ref{Fig5}-\ref{Fig6} show the phase diagrams for France and for Italy, respectively. The diagrams are built in terms of ensemble averages of number of infections per day $I(t)$ versus the average value of $R_0$ after the confinement (panels a), and the errors (represented as standard deviation of the average $I(t)$ over the 30 realisations) are shown in panels b. First we note that despite some small differences in the delay of the COVID-19 second wave of infections peak, the diagrams are very similar. In order to avoid a second wave, $R_0$ could fluctuate on values even slightly larger than one. Furthermore, for  $1.5<R_0<2$, the second wave is delayed in Autumn or Winter 2020/2021 months. The uncertainty follows the same behavior as the average and it peaks when the number of daily infections is maximum. This means that the ability to control the outcome of the epidemics is highly reduced if $R_0$ is too high.

\section{Discussion}
\label{sec:discussion}

France and Italy have faced a long phase of lockdown with severe restrictions in mobility and social contacts. They have managed to reduce the number of daily COVID-19 infections drastically and released almost simultaneously lockdown measures. This paper  addresses the possible future scenarios of COVID-19 infections in those countries by using one of the simplest possible model capable to reproduce the first wave of infections and to take into account uncertainties, namely a stochastic SEIR model with fluctuating parameters.\\

We have first verified that the model is capable to reproduce the behavior of the first wave of infections and provide an estimate of COVID-19 prevalence that is coherent with clinical tests and other studies. The introduction of stochasticity accounts for the large uncertainties in both the initial conditions as well as the fluctuations in the basic reproduction number $R_0$ originating from changes in virus characteristics, mobility or misapplication in confinement measures. 30 realisations of the model have been produced and they show very different COVID-19 prevalence after the first wave. The range goes from thousands of infected to tens of millions of infections in both countries. Average values are compatible with those found in other studies~\cite{Salje20,flaxman2020report}. \\

Then, we have modelled future epidemics scenarios by choosing specific fluctuating behaviors for $R_0$ and performing again, 30 realisations of the stochastic SEIR model. Despite the very large uncertainties, distinct scenarios clearly appear from the noise. 
In particular, they suggest that a second wave  can be avoided  even with $R_0$ values slightly larger than one. This means that actual distancing measures which include the use of surgical masks, the reduction in mobility and the active contact tracking can be effective in avoiding a second peak of infections without the need of imposing further strict lockdown measures. The analysis of phase diagrams show that there is a sharp transition between observing or not a second wave of infections when the value of $R_0$ is close to 1.5. Moreover, the models show that the higher $R_0$, the lower the ability to control the number of infections in the epidemics. 

This model has also evident deficiencies in representing the COVID-19 infections. First of all, the choice of the initial conditions is conditioned by our ignorance on the diffusion of the virus in France and Italy in December 2019. Furthermore, we are unable to verify on an extensive dataset the outcome of the first wave: on one side antibodies blood tests have still a lower reliability~\cite{long2020antibody} and on the other they have not been applied on   an extensive number of individuals to get reliable estimates. On top of the data-driven limitations, we have those introduced by the use of compartment models, as there are geographic, social and age differences in the spread of the COVID-19 disease in both countries~\cite{di2020expected}. Furthermore, we also assume that fluctuations on the parameters of the SEIR model are Gaussian, although there are good reasons to think that they should be heavy tailed distributions~\cite{maleki2020time,liu2020secondary}. We would like to remark however that, to overcome these limitations, one would need to fit more complex models and introduce additional parameters which can, at the present stage, barely inferred by the data.\\

Our choice to stick the stochastic SEIR model is indeed driven by its simplicity and the possibility of modeling realistic the uncertainties with the stochastic fluctuations instead of adding new parameters whose inference may affect the results. This study can be applied to other countries, and this is why we publish along the code of our analysis alongside with the paper. To date, Northern Europe, UK, US and other American countries are still facing the first wave of infections, so that future scenarios cannot be devised with the same clarity as those outlined in this study for France and Italy.

\section{Acknowledgments}
 DF acknowledges All the London Mathematical Laboratory fellows, B Dubrulle, F Pons, N Bartolo, F Daviaud, P Yiou, M Kagayema, S Fromang and G Ramstein for useful discussions.

\section{Data Availability}
The data that support the findings of this study are openly available in~\url{https://systems.jhu.edu/research/public-health/ncov/},
maintained by  Johns Hopkins University Center for Systems Science.

\section{Appendix A: numerical code}
\begin{verbatim}
% This appendix contains the MATLAB code used to perform
% the analysis contained in the paper via a stochasitc
% SEIR model

%% PARAMETER DEFINITIONS
%tmax: number of day of integrations
tmax=500;
%nrel: number of realisations of the model
nrel=30;
%tconf: lockdown day
tconf=50
%tconf2: lockdown release
tconf2=100

%% LOOP ON DIFFERENT VALUES OF LAMBDA, INFECTION RATE
for la=1:50
    lambdaconf=0.25;
    lambdares=la.*0.02;

    %% LOOP ON REALIZATIONS
    for rel=1:nrel
        S=zeros(1,tmax);
        E=zeros(1,tmax);
        I=zeros(1,tmax);
        R=zeros(1,tmax);
        C=zeros(1,tmax);
        lambda=zeros(1,tmax);
        %S Susceptible individuals (France population)
        S(1)=67000000;
        %I Infected individuals
        I(1)=585;
        % Recovered
        R(1)=0;
        % Inital time
        T(1)=0;
        % Cumulative infections
        C(1)=0;
        % alpha is the inverse of the incubation period (1/t_incubation)
        alpha0=0.27;
        % R0 is equal to 2.68
        R0=2.68;
        % gamma is the inverse of the mean infectious period
        gamma0=lambda0./R0;
        % uncertainty in gamma and lambda
        coeff_gamma=0.5;
        coeff_lambda=0.005;

            %% LOOP ON TIME, INTEGRATION OF SEIR MODELS
            for t=1:1:tmax
            %gamma=1/Tr where Tr is the recovery time  (2 weeks)
            %Stochastic gamma
            gamma=gamma0+gamma0./5*randn;
            %Change lambda for confinement
            if t==tconf
            lambda0=lambdaconf;
            end
            if t==tconf2
            lambda0=lambdares;
            end
            %Stochastic lambda
            lambda(t+1)=(lambda0+lambda0./5*randn)./S(1);
            %Stochastic alpha
            alpha=alpha0+alpha0./5*randn;
            %Computation of R0
            R0(t+1)=lambda(t+1)./gamma0;
            %Iteration of the model
            T(t+1)=t;
            S(t+1)=S(t)-(lambda(t+1)*S(t)*I(t));
            E(t+1)=E(t)+(lambda(t+1)*S(t)*I(t))-alpha*E(t);
            I(t+1)=I(t) +alpha*E(t)  -gamma*I(t);
            R(t+1)=R(t)+(gamma*I(t));
            %cumulative infected
            C(t+1)=gamma0.*sum(I);
            %Variables for different realisations
            Irel(rel,t+1)=I(t+1);
            lambdarel(rel,t+1)=lambda(t+1);
            end
            
        end
        
%% AVERAGING OVER DIFFERENT REALIZATIONS
lambdamoy(la,:)=mean(lambdarel,1);
Imoy(la,:)=mean(Irel,1);
Istd(la,:)=std(Irel,1);
lambdavec(la)=lambdares;
R0moy(la,:)=lambdamoy(la,:)./gamma0.*S(1);

end

\end{verbatim}

\bibliography{COVID}

\newpage

\end{document}